\documentclass{article}
\usepackage{spconf,amsmath,graphicx,hyperref}
\usepackage{amssymb}
\usepackage{amsfonts}
\usepackage{amsthm}
\usepackage{cite}
\usepackage{algorithm}
\usepackage{algpseudocode}
\usepackage{placeins}
\usepackage{capt-of}
\usepackage{comment}
\usepackage{bbm}
\usepackage{todonotes}
\usepackage{orcidlink}
\usepackage{svg}
\usepackage{subcaption}
\usepackage{enumitem}

\newcommand{\Mod}[1]{\mathcal{M}_{\lambda}\!\left(#1\right)}
\newcommand{\sinc}{\operatorname{sinc}}
\newcommand{\OF}{\mathrm{OF}}
\newcommand{\E}{\mathbb{E}}
\newcommand{\bfa}{\mathbf{a}}

\newcommand{\bfI}{\mathbf{I}}
\newcommand{\bfw}{\mathbf{w}}

\newcommand{\bfr}{\mathbf{r}}
\newcommand{\bfq}{\mathbf{q}}

\newcommand{\SigY}{\boldsymbol{\Sigma}_{Y}}
\newcommand{\SigW}{\boldsymbol{\Sigma}_{w}}

\newcommand{\bfx}{\mathbf{x}}
\newcommand{\bfn}{\mathbf{n}}
\newcommand{\bfm}{\mathbf{m}}
\newcommand{\bfom}{\boldsymbol{\omega}}
\newcommand{\SigTot}{\boldsymbol{\Sigma}_{T}}

\title{Fold First, Detect Directly: Communication Symbol Detection Without Unfolding for Low-Bitrate Modulo-ADCs}

\name{Krunal Vaghela~\orcidlink{0009-0001-0546-2859},
      Kumar Appaiah~\orcidlink{0000-0002-3149-4416}, and
      Satish~Mulleti~\orcidlink{0000-0002-3995-9070}}
\address{Department of Electrical Engineering, Indian Institute of Technology Bombay, Mumbai, India, 400076.\\
Emails: {krunalvaghela1262@gmail.com}, {akumar@ee.iitb.ac.in}, {mulleti.satish@gmail.com}}

\begin{document}
\ninept
\maketitle

\setcounter{topnumber}{4}
\setcounter{bottomnumber}{2}
\setcounter{totalnumber}{6}
\renewcommand{\topfraction}{0.95}
\renewcommand{\bottomfraction}{0.8}
\renewcommand{\textfraction}{0.05}
\renewcommand{\floatpagefraction}{0.75}
\raggedbottom

\begin{abstract}
Modulo-folding analog-to-digital converters (MF-ADCs) enable low-dynamic-range quantizers to sample high-amplitude signals without clipping. However, downstream processing traditionally relies on waveform unfolding algorithms, which require high oversampling rates and are highly sensitive to noise. For communication receivers, where the goal is symbol detection rather than signal reconstruction, unfolding is a redundant intermediate step. In this paper, we propose an unfolding-free maximum-likelihood symbol-detection framework for oversampled MF-ADCs in the presence of joint channel and quantization noise. By leveraging modulo wrap cancellation, we derive an exact, single-term Mahalanobis-distance metric that operates directly on folded observations. To handle long sequences, we introduce a parallelized block search algorithm that reduces computational complexity to scale linearly with sequence length. Simulations show our detector significantly outperforms existing unfolding baselines and approaches unclipped conventional ADC performance.
\end{abstract}


\section{Introduction}
\label{sec:intro}

The dynamic range (DR) of an analog-to-digital converter (ADC) plays a critical role in high-speed digital communications and signal acquisition. While an ADC's DR must exceed the peak amplitude of the input signal to prevent clipping, conventional high-DR ADCs require high bit resolution to suppress quantization noise, leading to prohibitive power consumption. To relax these dynamic range constraints, Bhandari \textit{et al.} introduced the \emph{unlimited sampling} framework \cite{bhandari2017unlimited, bhandari2020unlimited}. In this framework, the continuous-time signal is folded via a modulo operation prior to sampling, restricting its amplitude range within $[-\lambda,\lambda]$ without information loss. Consequently, a low-DR ADC with few bits can acquire high-DR signals without clipping. Modulo-folding ADCs (MF-ADCs) have since been successfully applied to sparse and finite-rate-of-innovation recovery \cite{uls_sparse, uls_fri, bhandari_ussr}, sinusoidal mixtures \cite{uls_sinmix}, multi-dimensional and graph signals \cite{uls_md, uls_graph2}, direction-of-arrival estimation \cite{uls_doa}, the modulo Radon transform \cite{uls_radon}, array processing \cite{fernandez2022computational}, bandpass demodulation \cite{uls_bandpass}, neural recording \cite{chen_digassist}, spiked-covariance estimation \cite{romanov2022spiked}, radar \cite{feuillen_radar22}, and massive-MIMO receivers \cite{uls_mimo}, alongside dedicated hardware implementations \cite{mulleti2023hardware, mulleti2025power}.

While MF-ADCs generate folded, low-DR samples, most downstream signal processing algorithms rely on recovering the original unfolded sequence. To this end, classical modulo recovery algorithms—ranging from higher-order differences \cite{bhandari2017unlimited, bhandari2020unlimited} and linear prediction (referred as CBHF)\cite{romanov2019above} to Fourier-Prony recovery \cite{bhandari2022fourierprony}, iterative sieving (ITER-SIS) \cite{guo2023itersis}, residual recovery (termed as B2R2) \cite{azar2023b2r2}, hysteresis \cite{florescu_hyst, florescu_hyst22}, and local averaging \cite{florescu_local}—have been developed. However, these methods suffer from major limitations: they require substantial oversampling to resolve non-linear wrap discontinuities, exhibit extreme sensitivity to post-folding quantization noise, and reconstruct the full analog waveform, which represents an unnecessary intermediate step for downstream inference tasks.

In digital communications, the ultimate goal is symbol detection rather than complete signal reconstruction. Addressing this, Mulleti \textit{et al.} \cite{mulleti2025low} recently demonstrated that for linearly modulated signals, symbol recovery can be achieved directly at the symbol rate without waveform unfolding. By designing injective constellations with respect to the modulo boundary, they showed that the oversampling overhead of traditional recovery algorithms can be eliminated. However, operating strictly at the Nyquist rate renders symbol-rate detection highly sensitive to noise, causing symbol error rates (SER) to degrade severely in low-to-moderate signal-to-noise ratio (SNR) regimes.

In this paper, we bridge this performance gap by establishing an unfolding-free Maximum Likelihood (ML) detection framework for oversampled MF-ADC receivers. Rather than treating oversampling as an unwrapping tool, we exploit the spatio-temporal noise correlation introduced by oversampling to perform direct detection on folded, quantized observations. The main contributions of this paper are summarized as follows:
\begin{itemize}
    \item We formulate the optimal ML sequence detector for oversampled MF-ADCs under joint additive channel noise and uniform quantization noise. By exploiting modulo wrap cancellation properties, we derive an exact, single-term Mahalanobis-distance (MHBL) metric operating directly on the folded observations.
    \item To bypass the exponential computational complexity of exhaustive sequence detection over long transmission horizons, we introduce a parallelized, block-decomposed search algorithm (Algorithm~\ref{alg:block}) that achieves linear execution complexity $\mathcal{O}(N_s)$ while preserving near-optimal SER.
    \item Numerical evaluations show that the proposed MHBL detector significantly outperforms state-of-the-art waveform unfolding algorithms (ITER-SIS \cite{guo2023itersis}, B2R2\cite{azar2023b2r2}, and CBHF \cite{romanov2019above}) and approaches the performance bound of an unclipped full-scale conventional ADC oracle.
\end{itemize}

The remainder of this paper is organized as follows. Section~\ref{sec:problem_formulation} introduces the system model. Section~\ref{sec:proposed} derives the Mahalanobis ML detector and block search algorithm. Section~\ref{sec:results} presents experimental results, followed by conclusions.



\begin{figure*}[t]
\centering
\includegraphics[width=\textwidth]{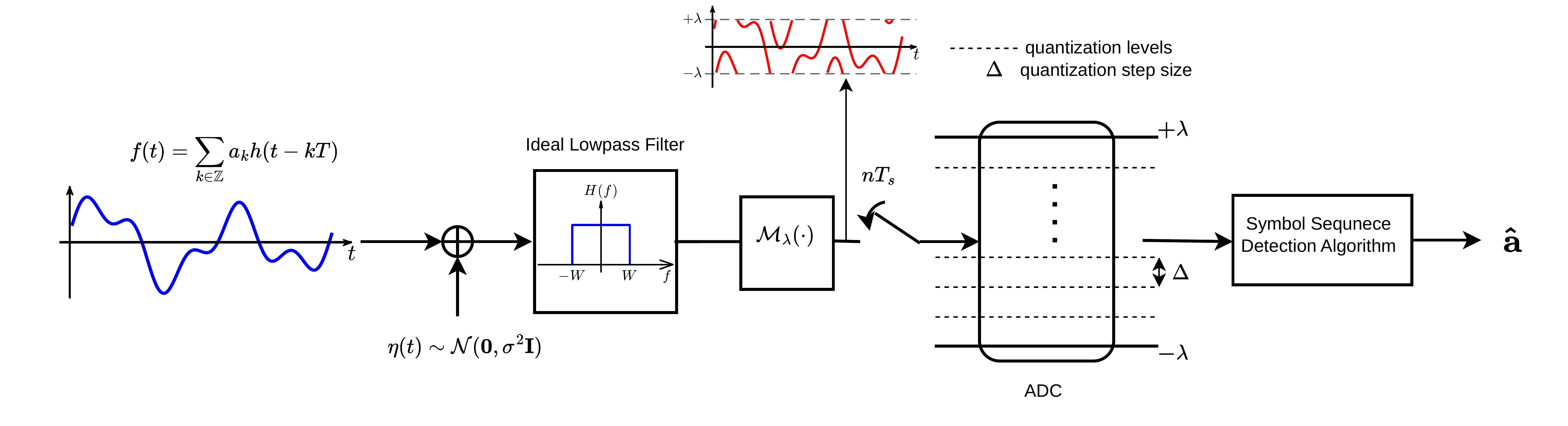}
\caption{Receiver pipeline: the ideal lowpass filter output is folded and quantized by the modulo-ADC ($2^R$ levels, step size $\Delta$), and the samples go directly to the symbol sequence detection algorithm.}
\label{fig:pipeline}
\end{figure*}
\begin{figure}[tb]
\centering
\includegraphics[width= 3 in]{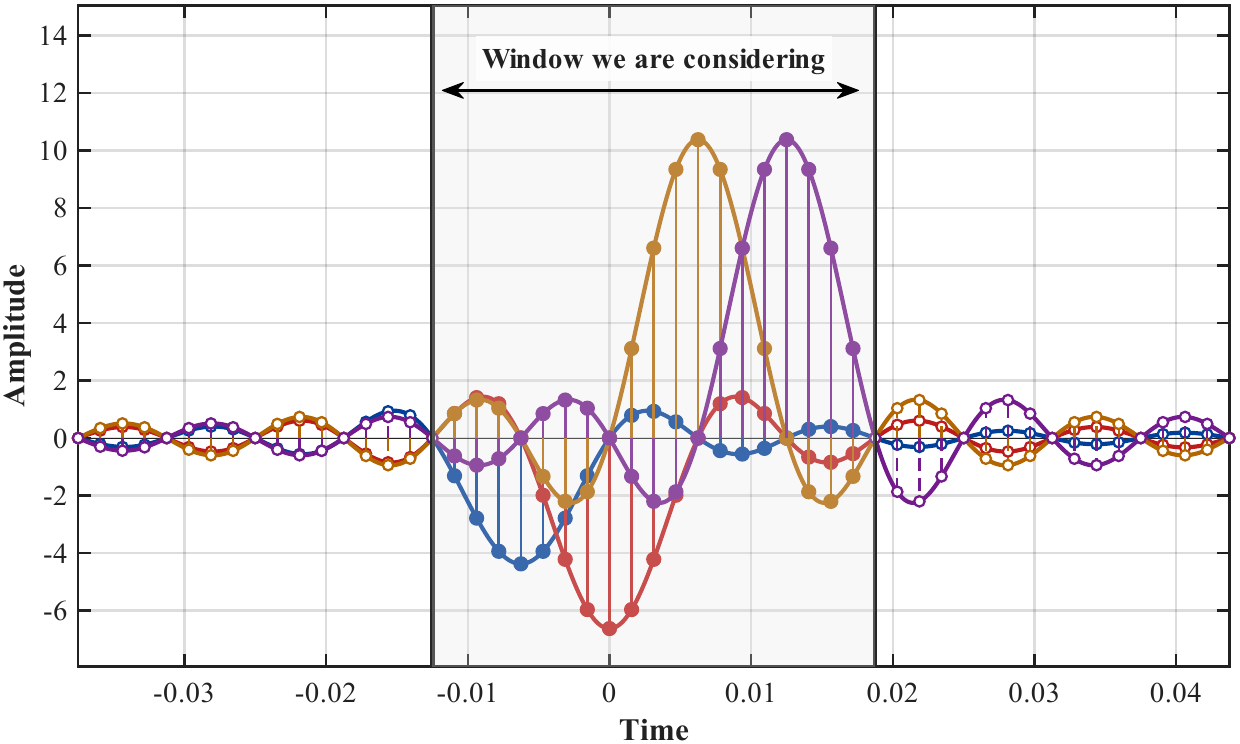}
\caption{Time-domain observation window bounded by the main lobe peaks of the boundary sinc pulses ($N_{s} = 4$ symbols).}
\label{fig:window}
\end{figure}

\section{System Model}
\label{sec:problem_formulation}

Consider a linearly modulated signal $f(t) = \sum_{m=1}^{N_s} a_m g(t - mT)$, where the symbols $a_m$ are drawn from a known alphabet $\mathcal{S} = \{s_1, \dots, s_M\}$. We assume that $g(t) = \operatorname{sinc}(2Wt)$ is bandlimited to $[-W, W]$ with symbol duration $T = 1/(2W)$. Consequently, the transmitted symbols correspond directly to the Nyquist samples $f(nT) = a_n$.

In practice, the communication signal is corrupted by an additive noise process $\eta(t)$ that is not necessarily bandlimited to $[-W, W]$. To minimize effective noise, the received signal $f(t) + \eta(t)$ is passed through the receiver's front-end lowpass filter with passband $[-W, W]$. For a general pulse shape $g(t)$, optimal detection requires a matched filter $g^*(T-t)$; however, because $g(t) = \operatorname{sinc}(2Wt)$, the ideal lowpass filter serves as its matched filter. This filtering leaves $f(t)$ completely unchanged, while reshaping the wideband noise $\eta(t)$ into a correlated, bandlimited Gaussian process $w(t)$ with autocorrelation $R_w(\tau) = \sigma_{\mathrm{lpf}}^2 \operatorname{sinc}(2W\tau)$, where $\sigma_{\mathrm{lpf}}^2$ denotes the resulting in-band noise power.

To avoid receiver clipping without requiring a high-dynamic-range ADC, the filtered signal $r(t) = f(t) + w(t)$ is passed through a central modulo operator $\mathcal{M}_\lambda(x) \triangleq \operatorname{mod}(x+\lambda, 2\lambda) - \lambda \in [-\lambda, \lambda)$ prior to sampling, yielding the dynamic-range-bounded signal $r_\lambda(t) = \mathcal{M}_\lambda(r(t)) \in [-\lambda, \lambda)$.

Unlike conventional ADCs that sample at the Nyquist rate, modulo-ADCs typically measure signals above the Nyquist rate to counter the non-linear folding effects and noise. Let the sampling rate be $F_s = \frac{1}{T_s} = \mathrm{OF} \times 2W$, where $\mathrm{OF}$ denotes the oversampling factor. Over a finite observation window of $N$ samples (Fig.~\ref{fig:window}) quantized using $R$ bits, the $N$-length folded sample vector $\mathbf{y}_\lambda$, unfolded sample vector $\mathbf{f}$, and symbol vector $\mathbf{a} \in \mathcal{S}^{N_s}$ are related as:
\begin{equation}
    \mathbf{y}_\lambda = \mathcal{M}_\lambda(\mathbf{f} + \mathbf{w}) + \mathbf{q} = \mathcal{M}_\lambda(\mathbf{H}\mathbf{a} + \mathbf{w}) + \mathbf{q},
    \label{eq:quant_model}
\end{equation}
where $\mathbf{w} \sim \mathcal{N}(\mathbf{0}, \mathbf{\Sigma}_w)$ is the zero-mean continuous noise vector with diagonal entries equal to $\sigma_{\mathrm{lpf}}^2$. Equivalently, \eqref{eq:quant_model} can be decomposed as:
\begin{equation}
    \mathbf{y}_\lambda = \mathbf{H}\mathbf{a} + \mathbf{w} + 2\lambda\mathbf{z} + \mathbf{q},
    \label{eq:modulo_observation_model}
\end{equation}
where $\mathbf{z} \in \mathbb{Z}^N$ is an element-wise integer folding vector. The vector $\mathbf{q}$ represents uniform quantization noise with entries distributed over $\left[-\frac{\lambda}{2^R}, \frac{\lambda}{2^R}\right]$, yielding a quantization noise variance of $\sigma_q^2 = \frac{\lambda^2}{3 \cdot 2^{2R}} = \frac{\Delta^2}{12}$ with step size $\Delta = \frac{\lambda}{2^{R-1}}$. 

The overall receiver pipeline is illustrated in Fig.~\ref{fig:pipeline}.In this oversampled regime, the interpolation matrix $\mathbf{H} \in \mathbb{R}^{N \times N_s}$ has entries $[\mathbf{H}]_{n,m} = \operatorname{sinc}(2W(nT_s - mT))$. At $\mathrm{OF} = 1$, $\mathbf{H} = \mathbf{I}$, recovering the symbol-rate setting of \cite{mulleti2025low}.

Our primary objective is to estimate the symbol vector $\mathbf{a}$ directly from the folded observations $\mathbf{y}_\lambda$, bypassing unfolding entirely.


\section{The ML Problem and Wrap Cancellation}
\label{sec:proposed}

A standard ML framework seeks the candidate symbol vector $\hat{\mathbf{a}}$ from the set of all possible symbol sequences $\mathcal{S}^{N_s}$ that maximizes the probability of observing measurements $\mathbf{y}_\lambda$:
\begin{equation}
    \hat{\mathbf{a}}_{\mathrm{ML}} = \arg\max_{\hat{\mathbf{a}} \in \mathcal{S}^{N_s}} \; p(\mathbf{y}_\lambda \mid \hat{\mathbf{a}}),
    \label{eq:ml}
\end{equation}
where $p(\mathbf{y}_\lambda \mid \hat{\mathbf{a}})$ denotes the conditional probability density function (PDF) of $\mathbf{y}_\lambda$ given $\mathbf{a}$. 

To derive this conditional density, we initially assume: (a) zero quantization noise ($\mathbf{q} = \mathbf{0}$), and (b) a deterministic folding residue $\mathbf{z}$. Because $\mathbf{w}$ follows a zero-mean multivariate Gaussian distribution $\mathcal{N}(\mathbf{0}, \boldsymbol{\Sigma}_w)$, the conditional density for a fixed $\mathbf{z}$ is given by:
\begin{align}
 p(\mathbf{y}_\lambda & \mid \mathbf{a}, \mathbf{z}) = 
 \frac{1}{(2\pi)^{N/2}|\boldsymbol{\Sigma}_w|^{1/2}} \times
\\ \nonumber
&\exp \Big( -\frac{1}{2} (\mathbf{y}_\lambda - \mathbf{H}\mathbf{a} - 2\lambda\mathbf{z})^\top 
  \boldsymbol{\Sigma}_w^{-1} (\mathbf{y}_\lambda - \mathbf{H}\mathbf{a} - 2\lambda\mathbf{z}) \Big).
\end{align}
Because the receiver cannot directly observe which integer wrap vector $\mathbf{z} \in \mathbb{Z}^N$ occurred, $p(\mathbf{y}_\lambda \mid \mathbf{a}, \mathbf{z})$ must be summed over all possible integer combinations:
\begin{equation}
    p(\mathbf{y}_\lambda \mid \mathbf{a}) = \sum_{\mathbf{z} \in \mathbb{Z}^N} p(\mathbf{y}_\lambda \mid \mathbf{a}, \mathbf{z}).
    \label{eq:sum}
\end{equation}
Evaluating \eqref{eq:sum} scales exponentially with dimension $N$, rendering direct likelihood computation computationally intractable in high-dimensional settings \cite{chemmala2024blind}. However, under practical operating conditions where the dynamic range parameter $\lambda$ lies comfortably above the system noise floor, this infinite sum can be accurately approximated by a single dominant lattice point.

\subsection{Mahalanobis Distance Formulation}
\label{sec:likelihood_truncation}

To resolve the infinite sum in \eqref{eq:sum}, we define the modulo residual vector $\mathbf{r}(\mathbf{a}) \triangleq \mathcal{M}_\lambda(\mathbf{y}_\lambda - \mathbf{H}\mathbf{a})$. Expressing the difference vector in terms of $\mathbf{r}(\mathbf{a})$ yields:
\begin{equation}
    \mathbf{y}_\lambda - \mathbf{H}\mathbf{a} = \mathbf{r}(\mathbf{a}) + 2\lambda \mathbf{z}^*,
    \label{eq:residual_decomposition}
\end{equation}
where $\mathbf{z}^* \in \mathbb{Z}^N$ is the unique integer shift centering the difference within $[-\lambda, \lambda)^N$. Substituting \eqref{eq:residual_decomposition} into \eqref{eq:sum} and re-indexing via $\mathbf{k} \triangleq \mathbf{z} - \mathbf{z}^*$ cancels the unknown shift $\mathbf{z}^*$ completely:
\begin{align}
p(\mathbf{y}_\lambda \mid \mathbf{a}) &= \sum_{\mathbf{k} \in \mathbb{Z}^N} \frac{1}{(2\pi)^{N/2}|\boldsymbol{\Sigma}_w|^{1/2}} \nonumber \\
&\times \exp \left( -\frac{1}{2} (\mathbf{r}(\mathbf{a}) - 2\lambda\mathbf{k})^\top \boldsymbol{\Sigma}_w^{-1} (\mathbf{r}(\mathbf{a}) - 2\lambda\mathbf{k}) \right).
\label{eq:simplified_pdf}
\end{align}

In operational regimes where $\lambda$ comfortably exceeds the peak noise ($\mathbf{w} \in [-\lambda,\lambda)^N$), applying $\mathcal{M}_\lambda(\cdot)$ to $\mathbf{y}_\lambda - \mathbf{H}\mathbf{a}$ simplifies directly:
\begin{equation}
    \mathbf{r}(\mathbf{a}) = \mathcal{M}_\lambda(\mathbf{y}_\lambda - \mathbf{H}\mathbf{a}) \overset{(a)}{=} \mathcal{M}_\lambda(\mathbf{w} + 2\lambda\mathbf{z}) \overset{(b)}{=} \mathcal{M}_\lambda(\mathbf{w}) \overset{(c)}{=} \mathbf{w},
    \label{eq:res_step3}
\end{equation}
where $\overset{(a)}{=}$ follows from \eqref{eq:modulo_observation_model}, $\overset{(b)}{=}$ holds because $2\lambda\mathbf{z}$ lies in the kernel of the modulo operator, and $\overset{(c)}{=}$ holds because $\mathbf{w} \in [-\lambda,\lambda)^N$. Comparing \eqref{eq:res_step3} with \eqref{eq:residual_decomposition} reveals that the true integer folding vector coincides with the shift vector ($\mathbf{z} = \mathbf{z}^*$, or $\mathbf{k} = \mathbf{0}$).

To quantify boundary excursions, consider the event $|w_n| > \lambda$, which occurs with probability $2Q(\lambda/\sigma_{\mathrm{lpf}})$ where $Q(x) = \frac{1}{2}\operatorname{erfc}(x/\sqrt{2})$. At $\lambda/\sigma_{\mathrm{lpf}} \ge 3$, this probability drops below $2Q(3) \approx 2.7 \times 10^{-3}$ per sample; an observation window of $N = 100$ samples contains only $0.27$ wrapped entries on average. Consequently, the unwrapped noise rarely crosses the folding boundary $[-\lambda, \lambda)^N$. As shown in Appendix~\ref{app:cov}, the second-order statistics of the wrapped noise $\mathcal{M}_\lambda(\mathbf{w})$ converge to those of the unwrapped noise $\mathbf{w}$ in this regime, rendering non-zero lattice terms ($\mathbf{k} \neq \mathbf{0}$) exponentially negligible.

Truncating \eqref{eq:simplified_pdf} to $\mathbf{k} = \mathbf{0}$ leads to:
\begin{equation}
    p(\mathbf{y}_\lambda \mid \mathbf{a}) \approx \frac{1}{(2\pi)^{N/2}|\boldsymbol{\Sigma}_w|^{1/2}} \exp \left( -\frac{1}{2} \mathbf{r}(\mathbf{a})^\top \boldsymbol{\Sigma}_w^{-1} \mathbf{r}(\mathbf{a}) \right).
    \label{eq:truncated_pdf}
\end{equation}

While \eqref{eq:truncated_pdf} accounts for continuous noise $\mathbf{w} \sim \mathcal{N}(\mathbf{0}, \boldsymbol{\Sigma}_w)$, physical receivers subject the folded signal to $R$-bit uniform quantization. The PDF of the uniform quantization error $\mathbf{q} \in \mathbb{R}^N$ over hypercube $\mathcal{H}_\Delta \triangleq [-\frac{\Delta}{2}, \frac{\Delta}{2}]^N$ is $f_{\mathbf{q}}(\mathbf{q}) = \frac{1}{\Delta^N} \mathbb{I}_{\mathcal{H}_\Delta}(\mathbf{q})$, where $\mathbb{I}_{(\cdot)}(\cdot)$ is the indicator function. The post-quantization likelihood $p_q(\mathbf{y}_\lambda \mid \mathbf{a})$ is given by $(\phi_{\boldsymbol{\Sigma}_w} * f_{\mathbf{q}})(\mathbf{r}(\mathbf{a}))$, where $\phi_{\boldsymbol{\Sigma}_w}(\cdot)$ is the right-hand side of  in \eqref{eq:truncated_pdf} and $f_{\mathbf{q}}(\cdot)$ is the quantization noise density. Because $p_q(\mathbf{y}_\lambda \mid \mathbf{a})$ lacks a closed-form solution for general covariance matrices $\boldsymbol{\Sigma}_w$, direct evaluation is intractable. However, when quantization noise variance is small relative to channel noise ($\sigma_q \ll \sigma_{\mathrm{lpf}}$), characteristic function expansion (Appendix~\ref{app:quant}) reveals that the effective joint distribution simplifies to a single Gaussian density with an inflated covariance matrix:
\begin{equation}
    \boldsymbol{\Sigma}_{T} = \boldsymbol{\Sigma}_w + \sigma_q^2 \mathbf{I}_N.
    \label{eq:SigmaTotal}
\end{equation}
Taking the negative log-likelihood reduces the post-quantization detector to a compact Mahalanobis-distance minimization over the modulo residual:
\begin{equation}
    \hat{\mathbf{a}} = \arg\min_{\mathbf{a} \in \mathcal{S}^{N_s}} \; \mathbf{r}(\mathbf{a})^\top \boldsymbol{\Sigma}_{T}^{-1} \mathbf{r}(\mathbf{a}).
    \label{eq:mahalanobis_ml_quantized}
\end{equation}
The metric in \eqref{eq:mahalanobis_ml_quantized} is our central theoretical result, enabling direct symbol detection from folded, quantized observations.

While evaluating \eqref{eq:mahalanobis_ml_quantized} is exact and effective for small symbol sequences ($N_s \le 10$), exhaustive search over all $|\mathcal{S}|^{N_s}$ candidates rapidly becomes computationally intractable for larger sequences. For instance, evaluating $N_s = 64$ symbols drawn from an $8$-ary constellation ($|\mathcal{S}| = 8$) requires testing $K = |\mathcal{S}|^{N_s} = 8^{64} \approx 6.27 \times 10^{57}$ hypotheses. To overcome this exponential search complexity, we propose a scalable, block-decomposed detection strategy next.

\subsection{Block-Decomposed Search for Long Symbol Sequences}
\label{subsec:block_search}

To break this exponential complexity, we exploit two physical decay characteristics of the system: sinc pulses $g(t) = \operatorname{sinc}(2Wt)$ decay as $\mathcal{O}(1/|t|)$, which limits inter-symbol interference (ISI) across long time spans, while the noise autocorrelation $R_w(\tau)$ vanishes for larger time separations $\tau$. By leveraging these decay properties, rather than processing all $N$ samples simultaneously to detect the $N_s$ symbols, we partition the measurement stream into smaller blocks and refine symbol estimates using an iterative, multi-pass procedure.

Consider the $k$-th pass evaluating the $j$-th block. Each block focuses on estimating $N_b$ core symbols, corresponding to an observation duration of $N_b \times \mathrm{OF}$ samples. To account for boundary ISI and noise correlation, a margin of $\eta \times \mathrm{OF}$ samples is appended to both sides of the block, yielding a total local observation window of $(N_b + 2\eta) \times \mathrm{OF}$ samples. Let $\mathcal{I}_j$ and $\mathcal{C}_j$ denote the sets of sample indices in the $j$-th block and local symbol indices, respectively.

Our detection approach incorporates symbol estimates from outside the current block obtained during the previous pass. We define two submatrices of $\mathbf{H}$: $\mathbf{H}(\mathcal{I}_j, :)$ containing all rows indexed by $\mathcal{I}_j$, and $\mathbf{H}(\mathcal{I}_j, \mathcal{C}_j)$ containing the submatrix corresponding to rows $\mathcal{I}_j$ and columns $\mathcal{C}_j$. The interference vector $\mathbf{d}_j^{(k-1)}$ caused by symbols external to the $j$-th block is calculated as:
\begin{equation}
    \mathbf{d}_j^{(k-1)} = \mathbf{H}(\mathcal{I}_j, :)\,\hat{\mathbf{a}}^{(k-1)} - \mathbf{H}(\mathcal{I}_j, \mathcal{C}_j)\,\hat{\mathbf{a}}^{(k-1)}_j,
    \label{eq:interference}
\end{equation}
where $\hat{\mathbf{a}}^{(k-1)}$ and $\hat{\mathbf{a}}^{(k-1)}_j$ denote the global and local symbol estimates from pass $k-1$, respectively. Subtracting $\mathbf{d}_j^{(k-1)}$ from the raw measurements yields the low-interference local observation vector $\mathbf{y}_j^{\text{eff}} = \mathbf{y}_\lambda(\mathcal{I}_j) - \mathbf{d}_j^{(k-1)}$.

Next, the updated local symbol vector $\hat{\mathbf{a}}_j^{(k)}$ is obtained as follows. Let $\mathbf{r}_j(\mathbf{\bar{a}}) \triangleq \mathcal{M}_\lambda\!\left(\mathbf{y}_j^{\text{eff}} - \mathbf{H}(\mathcal{I}_j, \mathcal{C}_j) \mathbf{\bar{a}}\right)$ denote the local modulo residual vector. Then, the local estimate $\hat{\mathbf{a}}_j^{(k)}$ is updated as:
\begin{equation}
    \hat{\mathbf{a}}_j^{(k)} = \arg\min_{\mathbf{\bar{a}} \in \mathcal{S}^{N_b}} \; \mathbf{r}_j(\mathbf{\bar{a}})^\top \boldsymbol{\Sigma}_{T,j}^{-1} \, \mathbf{r}_j(\mathbf{\bar{a}}).
    \label{eq:local_ml}
\end{equation}
where $\boldsymbol{\Sigma}_{T,j} = \boldsymbol{\Sigma}_w(\mathcal{I}_j, \mathcal{I}_j) + \sigma_q^2 \mathbf{I}_{N_j}$ is the local noise covariance matrix padded by the sample margin. After estimating $\hat{\mathbf{a}}_j^{(k)}$, edge symbols subject to severe boundary truncation errors are discarded. Block overlaps are configured such that these discarded edge symbols are properly centered and estimated in adjacent blocks. All blocks execute this process in parallel until the global vector converges or a maximum of $K$ passes is reached.

Assuming $P$ overlapping blocks, the overall detection process is summarized in Algorithm~\ref{alg:block}.

While exhaustive ML evaluation scales exponentially as \\
$\mathcal{O}(\mathrm{OF}^2 N_s^2 \, |\mathcal{S}|^{N_s})$, processing $P$ overlapping blocks reduces the computational complexity to $\mathcal{O}\!\left(K \cdot P \cdot \mathrm{OF}^2 (N_b + 2\eta)^2 \, |\mathcal{S}|^{N_b}\right).$
This cost increases \emph{linearly} with sequence length $N_s$ rather than exponentially. Mathematically, the rapid tail decay of sinc pulses ensures that symbols outside block $j$ exert negligible residual ISI on local symbols $\mathcal{C}_j$, allowing $\mathbf{H}(\mathcal{I}_j, \mathcal{C}_j)$ to capture dominant signal energy. Concurrently, the fast drop in noise autocorrelation $R_w(\tau)$ renders off-diagonal terms in $\boldsymbol{\Sigma}_T$ negligible beyond the sample margin $\mathcal{I}_j$, making $\boldsymbol{\Sigma}_{T,j}^{-1}$ a near-exact local precision matrix.

\begin{algorithm}[t]
\caption{Block-Mahalanobis Search}
\label{alg:block}
\begin{algorithmic}[1]
\Require Folded observation vector $\mathbf{y}_\lambda$; precomputed block parameters $\{(\mathcal{I}_j, \mathcal{C}_j, \mathbf{H}(\mathcal{I}_j, :), \mathbf{H}(\mathcal{I}_j, \mathcal{C}_j), \boldsymbol{\Sigma}_{T,j}^{-1})\}_{j=1}^{P}$; maximum passes $K$.
\Ensure Detected symbol vector $\hat{\mathbf{a}}$.
\State Initialize global symbol estimate $\hat{\mathbf{a}}^{(0)}$
\For{$k=1, \dots, K$}
    \For{$j=1, \dots, P$ \textbf{in parallel}}
        \State Compute interference $\mathbf{d}_j^{(k-1)}$ via \eqref{eq:interference}
        \State Form effective observation: $\mathbf{y}_j^{\text{eff}} \gets \mathbf{y}_\lambda(\mathcal{I}_j) - \mathbf{d}_j^{(k-1)}$
        \State \hspace{-.2 in}Get modulo residual: $\mathbf{r}_j(\mathbf{\bar{a}}) \gets \mathcal{M}_\lambda\big(\mathbf{y}_j^{\text{eff}} - \mathbf{H}(\mathcal{I}_j, \mathcal{C}_j)\mathbf{\bar{a}}\big)$
        \State \hspace{-.2 in}Local search: $\hat{\mathbf{a}}_j^{(k)} \gets \arg\min_{\mathbf{\bar{a}} \in \mathcal{S}^{N_b}} \; \mathbf{r}_j(\mathbf{\bar{a}})^\top \boldsymbol{\Sigma}_{T,j}^{-1} \, \mathbf{r}_j(\mathbf{\bar{a}})$
    \EndFor
    \State Discard edge symbols and merge core estimates from $\{\hat{\mathbf{a}}_j^{(k)}\}_{j=1}^P \to \hat{\mathbf{a}}^{(k)}$
    \If{$\hat{\mathbf{a}}^{(k)} = \hat{\mathbf{a}}^{(k-1)}$} \textbf{break} \EndIf
\EndFor
\State \Return $\hat{\mathbf{a}}^{(k)}$
\end{algorithmic}
\end{algorithm}


\section{Experimental Results}
\label{sec:results}

We evaluate performance using the SER, defined as $\mathrm{SER} = \frac{1}{N_s}\sum_{i=1}^{N_s}\mathbb{I}_{\{\hat{a}_i \neq a_i\}} \times 100\%$, averaged over $500$ Monte Carlo trials per operating point using the observation window and Nyquist spacing defined in Section~\ref{sec:problem_formulation}. Operating points are parameterized by the compression ratio $c/\lambda$, where $c = \max(\mathcal{S})$, and the pre-folding in-band SNR: $10\log_{10}\!\left(\frac{1}{|\mathcal{S}|\sigma_{\mathrm{lpf}}^2}\sum_{s\in\mathcal{S}} s^2\right)$ dB.

Because the modulo operator is many-to-one, we employ an injective 8-ary PAM constellation mapped to $\lambda$ as in \cite{mulleti2025low}. Setting $\sigma_{\mathrm{lpf}} = \lambda/3$ yields operating points $(c/\lambda, \mathrm{SNR}) \approx (10.4, 25.7\,\mathrm{dB})$, $(4.9, 19.1\,\mathrm{dB})$, and $(3.3, 16.7\,\mathrm{dB})$ for dynamic range parameters $\lambda = 1, 2, 3$, respectively. We evaluate our proposed MHBL detector against an unclipped full-scale conventional ADC oracle and three state-of-the-art waveform unfolding benchmarks: ITER-SIS \cite{guo2023itersis}, B2R2 \cite{azar2023b2r2}, and CBHF \cite{romanov2019above}.

\begin{figure}[tb]
\centering
\includegraphics[width=2in]{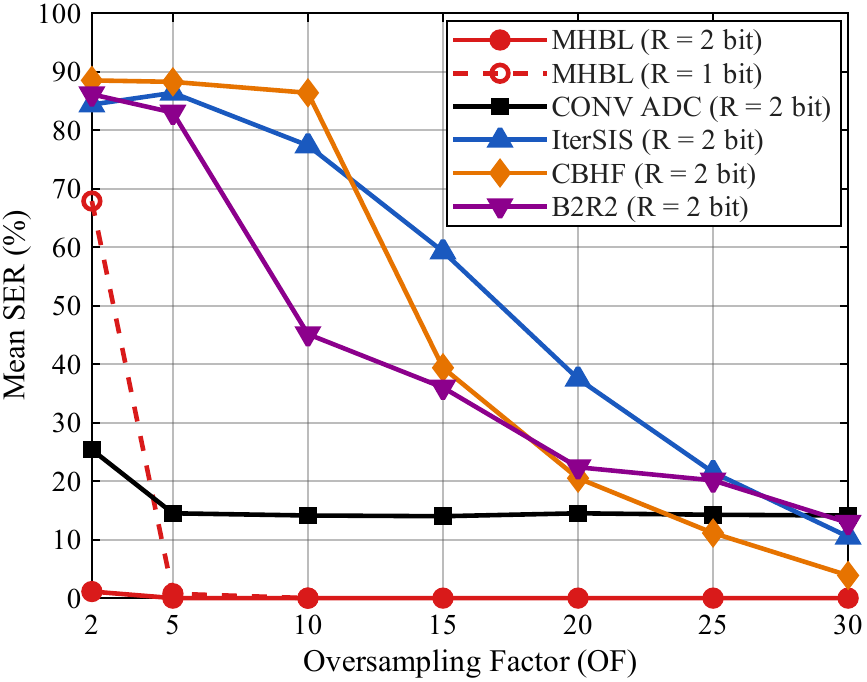}
\caption{SER versus oversampling factor ($\mathrm{OF}$) in the quantization-only regime ($\sigma_{\mathrm{lpf}}=0$, $N_{s}=4$, $c/\lambda=10.4$) for MHBL and conventional ADC oracle at $R \in \{1, 2\}$ bits.}
\label{fig:quant_only}
\end{figure}

\begin{figure}[tb]
\centering
\includegraphics[width=2in]{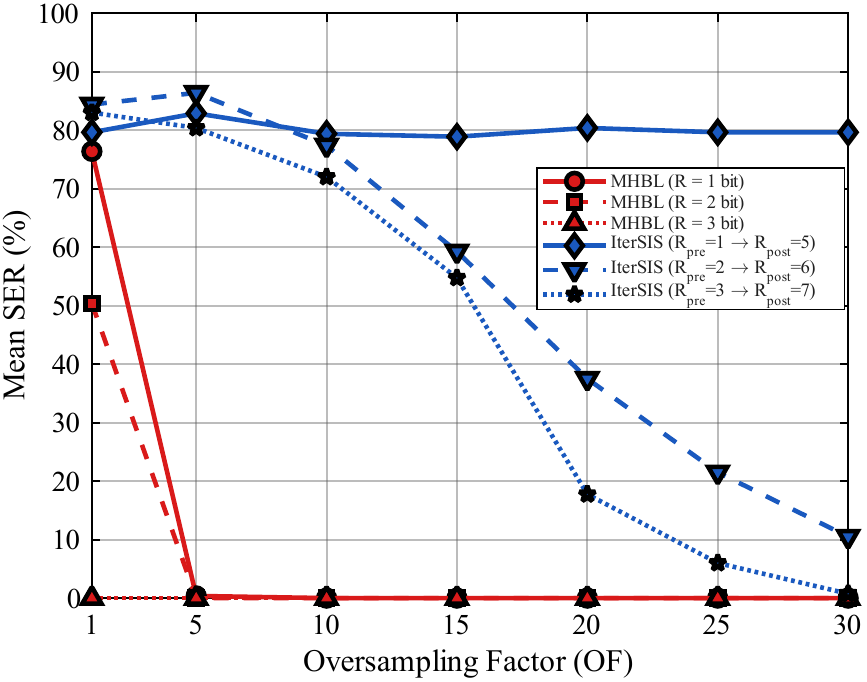}
\caption{Quantization-only regime: mean SER (\%) over resolution $R$ and $\OF$ for MHBL  and ITER-SIS}
\label{fig:heatmap_itersis}
\end{figure}

\begin{figure}[tb]
\centering
\includegraphics[width=2in]{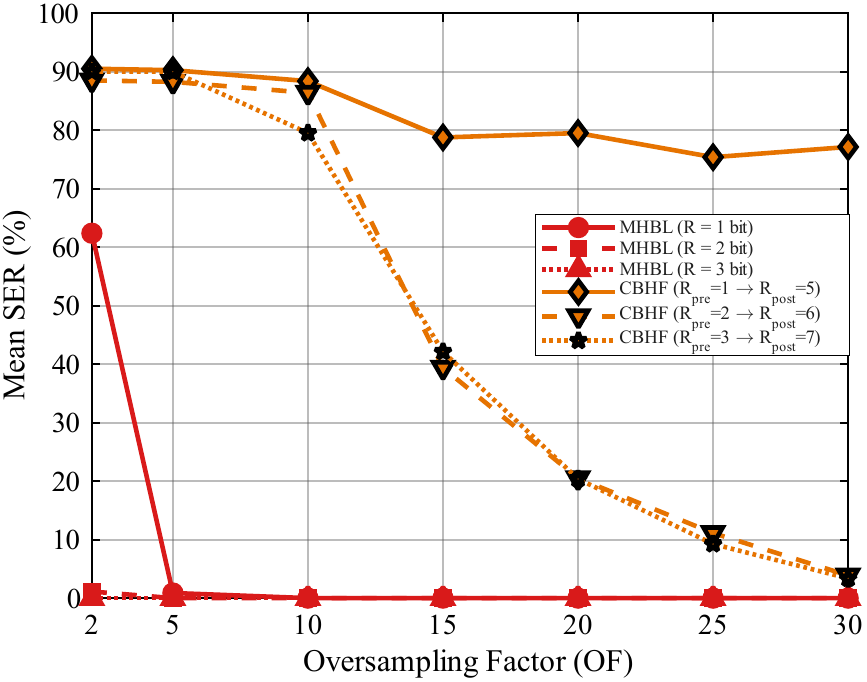}
\caption{Quantization-only regime: mean SER (\%) over resolution $R$ and $\OF$ for MHBL  and CBHF.}
\label{fig:heatmap_chbf}
\end{figure}

\begin{figure}[tb]
\centering
\includegraphics[width=2in]{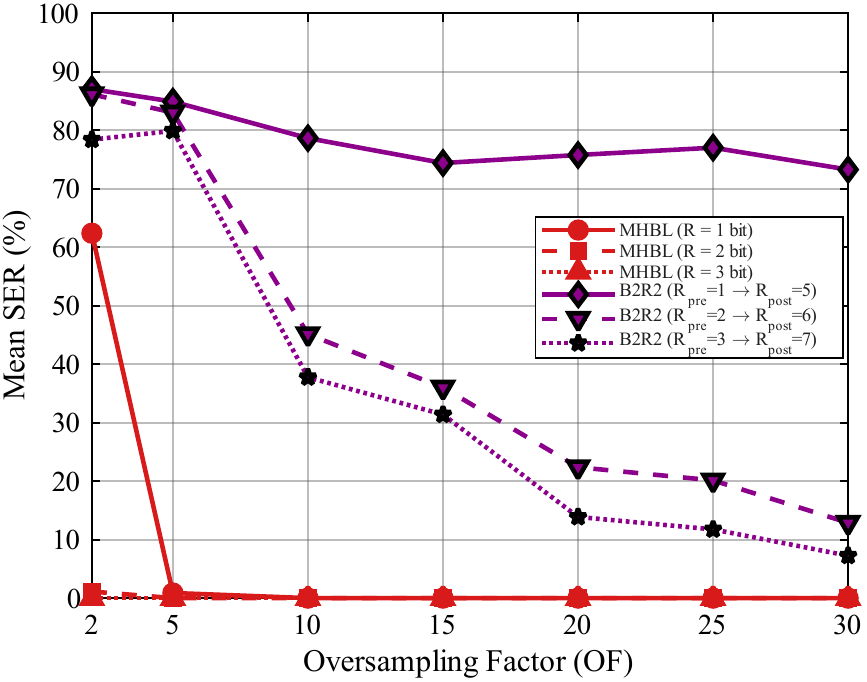}
\caption{Quantization-only regime: mean SER (\%) over resolution $R$ and $\OF$ for MHBL  and B2R2}
\label{fig:heatmap_b2r2}
\end{figure}

The first experiment evaluates performance strictly under $R$-bit quantization noise ($\sigma_{\mathrm{lpf}} = 0$, $c/\lambda = 10.4$). As shown in Fig.~\ref{fig:quant_only}, as $\mathrm{OF}$ increases from $2$ to $30$, the proposed MHBL detector outperforms all unfolding baselines by a wide margin. Unfolding methods struggle severely in this regime: ITER-SIS fails due to high compression ratios ($c/\lambda$) and thresholding errors magnified by coarse quantization, while B2R2 and CBHF suffers from spectral leakage caused by finite observation windows. Figs.~\ref{fig:heatmap_itersis}--\ref{fig:heatmap_b2r2} additionally compare MHBL separately with each of ITER-SIS, CBHF and B2R2 in the quantization-only regime for $R \in \{1,2,3\}$ bits.

\begin{figure}[tb]
\centering
\includegraphics[width=2in]{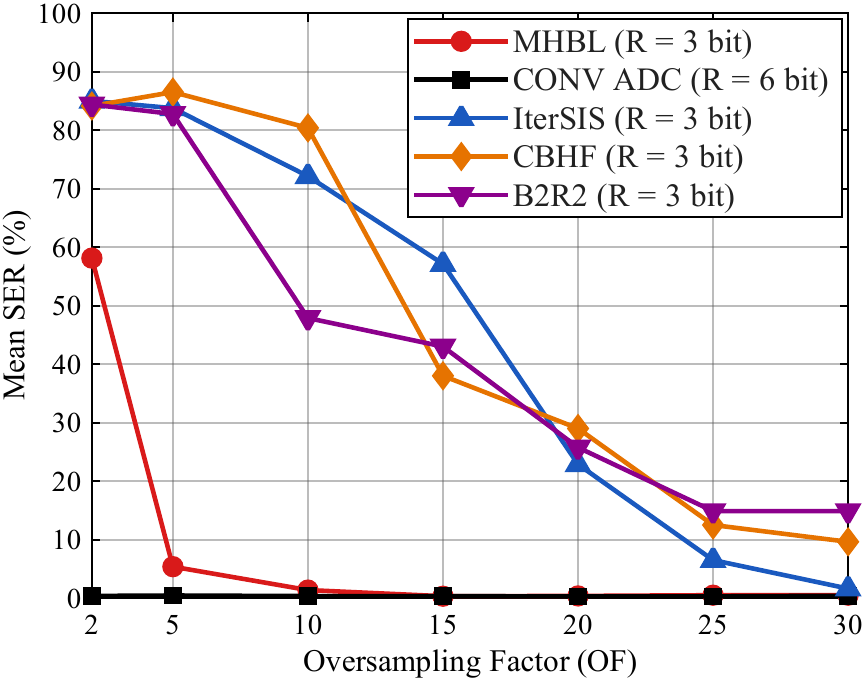}
\caption{SER vs.\ $\mathrm{OF}$ at $\lambda=1$ under joint noise ($N_s=4$, $c/\lambda\approx10.4$, $\lambda/\sigma_{\mathrm{lpf}}=3$, $\mathrm{SNR}\approx25.7\,\mathrm{dB}$).}
\label{fig:lambda1}
\end{figure}


\begin{figure}[tb]
\centering
\includegraphics[width=2 in]{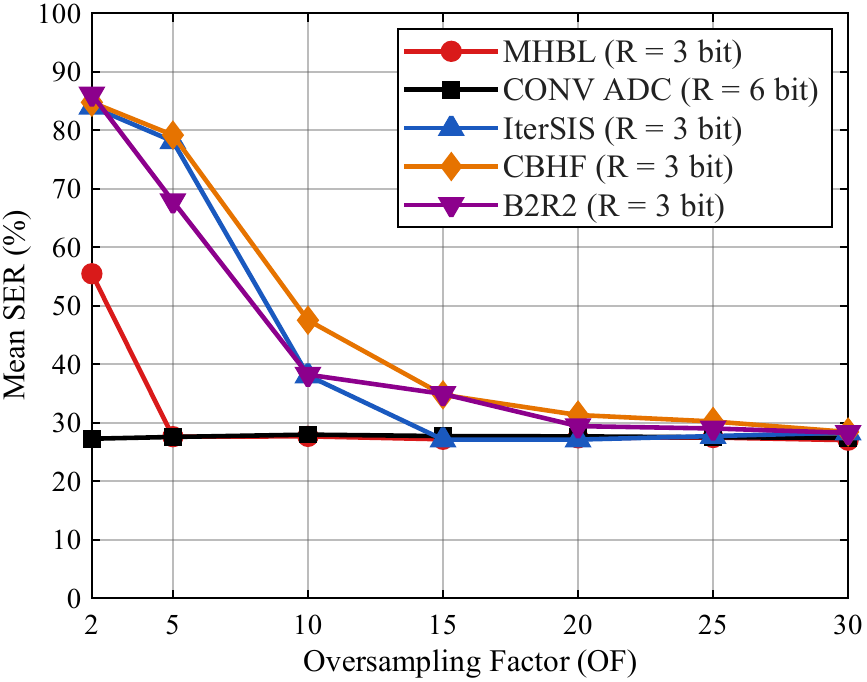}
\caption{SER vs.\ $\mathrm{OF}$ under joint noise at $\lambda=2$ ($N_s=4$, $c/\lambda\approx4.9$, $\lambda/\sigma_{\mathrm{lpf}}=3$, $\mathrm{SNR}\approx19.1\,\mathrm{dB}$).}
\label{fig:lambda2}
\end{figure}

\begin{figure}[tb]
\centering
\includegraphics[width=2 in]{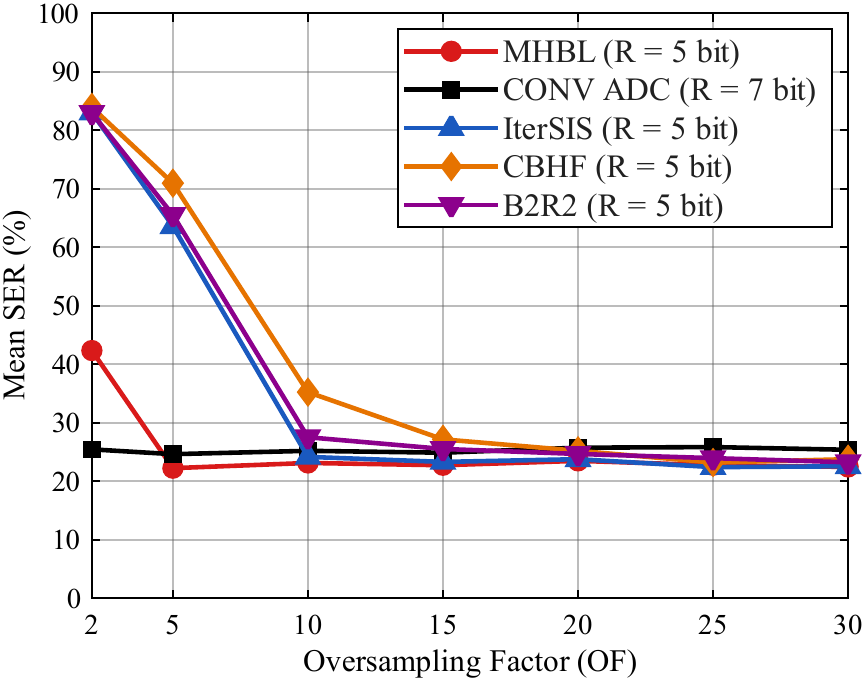}
\caption{SER vs.\ $\mathrm{OF}$ under joint noise at $\lambda=3$ ($N_s=4$, $c/\lambda\approx3.3$, $\lambda/\sigma_{\mathrm{lpf}}=3$, $\mathrm{SNR}\approx16.7\,\mathrm{dB}$).}
\label{fig:lambda3}
\end{figure}

\begin{figure}[tb]
\centering
\includegraphics[width=2 in]{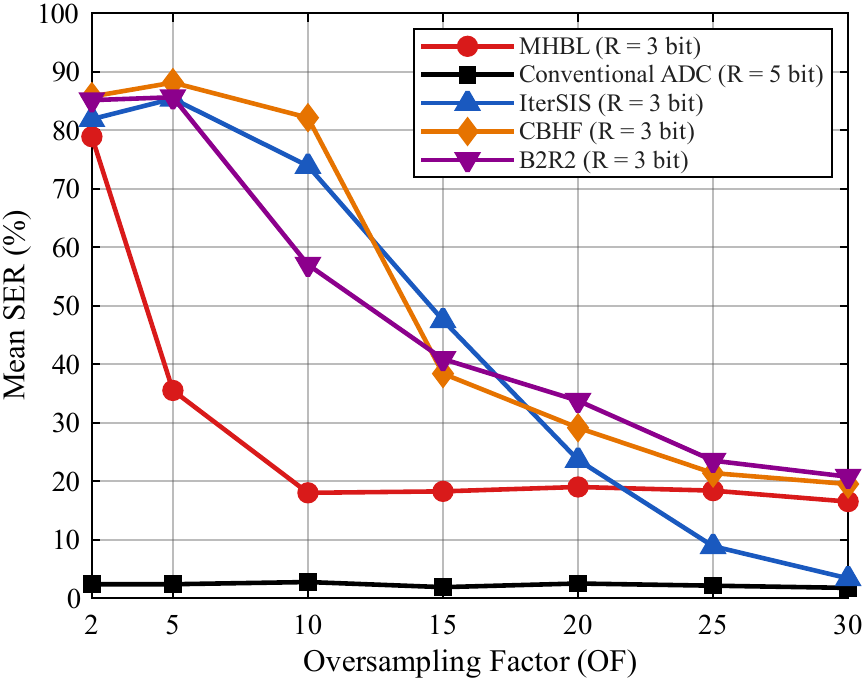}
\caption{Same comparison as Fig.~\ref{fig:lambda1}, at $\lambda=1$ ($c/\lambda\approx10.4$, $\lambda/\sigma_{\mathrm{lpf}}=2$, in-band SNR of the signal before modulo folding $\approx22.1$~dB).}
\label{fig:lambda1_2}
\end{figure}
Next, we evaluate performance under combined continuous channel noise and uniform quantization noise ($N_s = 4$, $\sigma_{\mathrm{lpf}} = \lambda/3$). Figures~\ref{fig:lambda1}--\ref{fig:lambda3} show the SER performance across varying $\lambda$ values. While the unclipped conventional ADC with high bit-depth sets the ideal performance lower bound, MHBL consistently achieves the lowest SER among all modulo-ADC approaches, particularly at low-to-moderate oversampling rates. For dynamic range parameter $\lambda=1$ and $\lambda/\sigma_{\mathrm{lpf}}=2$ (Fig.~\ref{fig:lambda1_2}), the SER of MHBL stays sufficiently above the conventional ADC baseline.

\begin{figure}[tb]
\centering
\includegraphics[width= 2in]{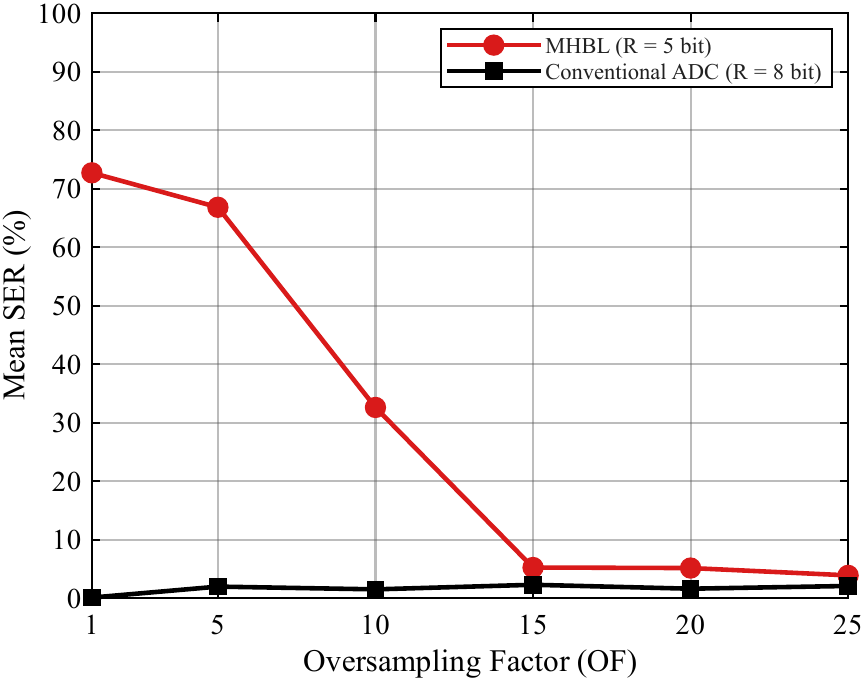}
\caption{SER vs.\ $\mathrm{OF}$ for a long sequence ($N_s=64$, $\lambda=1$, $c/\lambda\approx10.4$, $\lambda/\sigma_{\mathrm{lpf}}=3$, $\mathrm{SNR}\approx25.7\,\mathrm{dB}$).}
\label{fig:large_ns}
\end{figure}

Finally, we test long symbol sequences ($N_s = 64$) where exhaustive search is computationally impossible. Algorithm~\ref{alg:block} enables linear-complexity execution in $\mathcal{O}(N_s)$ time. As shown in Fig.~\ref{fig:large_ns}, a $5$-bit MHBL receiver rapidly approaches the performance of an $8$-bit conventional ADC oracle as $\mathrm{OF}$ increases. Unfolding algorithms could not be evaluated for $N_s = 64$ due to their prohibitive runtime and catastrophic error propagation over long time horizons.

\section{Conclusion}
\label{sec:conclusion}

We presented an unfolding-free Maximum Likelihood symbol detector for oversampled modulo-ADCs under joint channel and quantization noise. By leveraging integer wrap cancellation, we derived a compact Mahalanobis-distance metric that operates directly on folded observations. To enable detection over long sequences, we introduced a parallelized block-search algorithm with $\mathcal{O}(N_s)$ linear complexity. Simulations show that the proposed detector significantly outperforms existing unfolding baselines and approaches the unclipped performance of conventional ADCs for $\lambda/\sigma_{\mathrm{lpf}} \ge 3$. Extending this framework to lower SNR regimes and exploring lower-complexity local search heuristics remain key topics for future work.
\begin{figure*}[t]
\centering
\begin{minipage}[t]{0.32\textwidth}
    \centering
    \includegraphics[width=\linewidth]{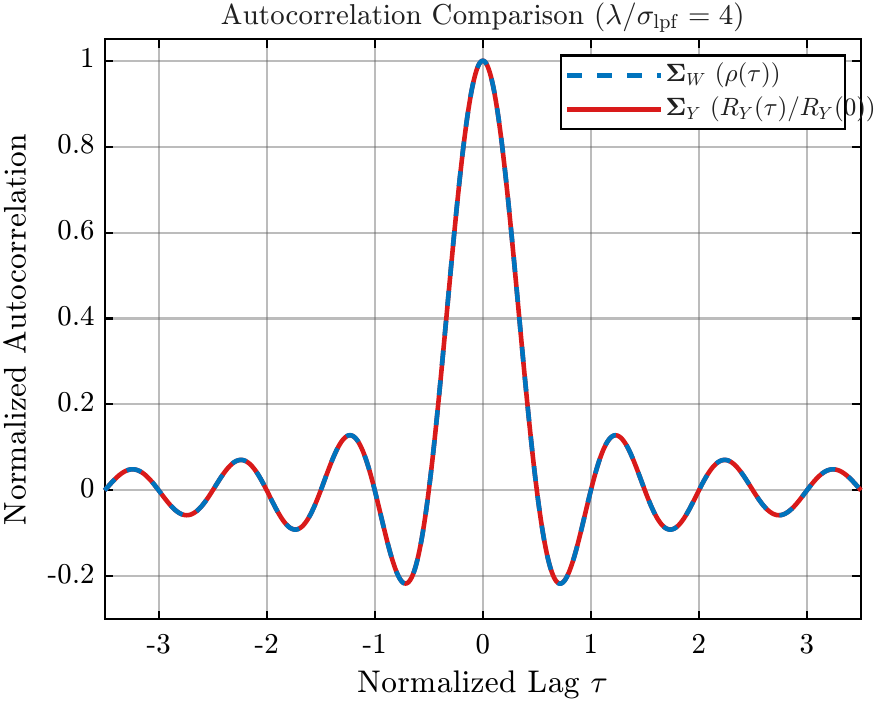}\\
    \small (a) $\lambda/\sigma_{\mathrm{lpf}}=4$
\end{minipage}\hfill
\begin{minipage}[t]{0.32\textwidth}
    \centering
    \includegraphics[width=\linewidth]{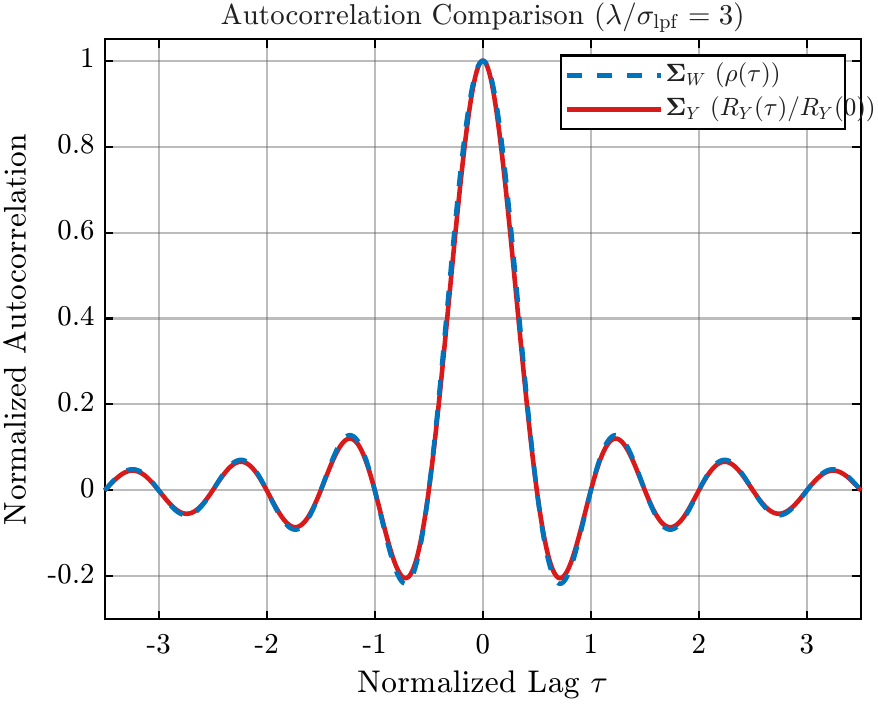}\\
    \small (b) $\lambda/\sigma_{\mathrm{lpf}}=3$
\end{minipage}\hfill
\begin{minipage}[t]{0.32\textwidth}
    \centering
    \includegraphics[width=\linewidth]{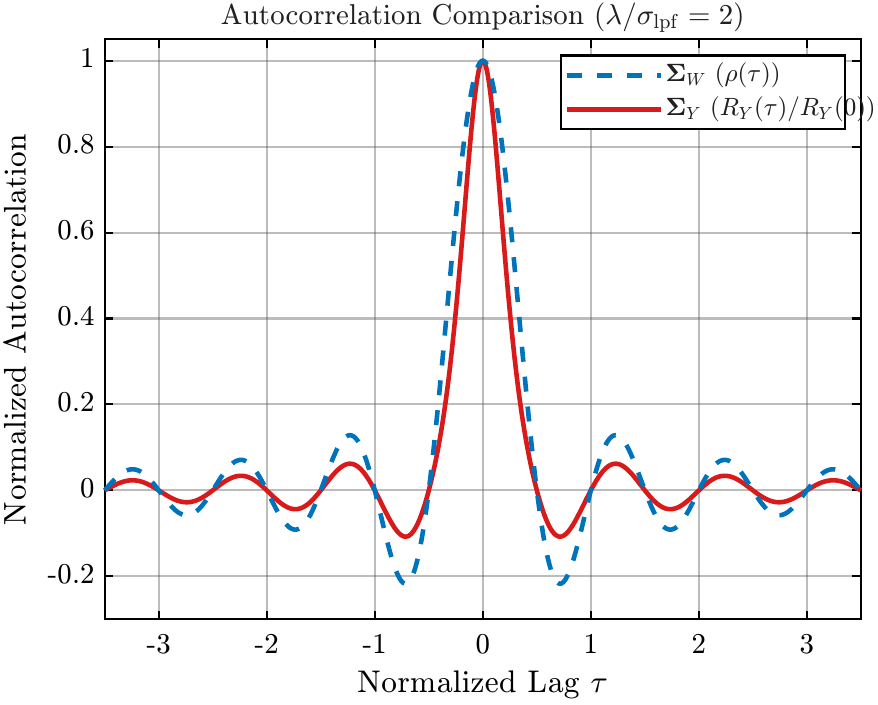}\\
    \small (c) $\lambda/\sigma_{\mathrm{lpf}}=2$
\end{minipage}
\caption{Normalized autocorrelation of the unfolded noise, $\rho(\tau)$, and of the folded noise, $R_Y(\tau)/R_Y(0)$, for different $\lambda/\sigma_{\mathrm{lpf}}$.}
\label{fig:autocorr}
\end{figure*}
\appendix
\section{Derivation of Folded Noise Autocorrelation and Analysis Relative to $\mathbf{R_w(\mathbf{\tau})}$}
\label{app:cov}

To support the performance analysis in Section~III, this appendix presents the analytical derivation of $R_Y(\tau)$ and evaluates how modulo folding alters the noise correlation profile relative to $R_w(\tau)$ across different  $\lambda/\sigma_{\mathrm{lpf}}$ ratios.

Since $\mathbf{w}$ is correlated through $R_w(\tau)=\sigma_{\mathrm{lpf}}^2\rho(\tau)$, where $\rho(\tau)=\mathrm{sinc}(2W\tau)$, we need the joint second-order statistics of $\mathcal{M}(\mathbf{w})$, i.e., the matrix
\begin{align}
    \boldsymbol{\Sigma}_Y &\triangleq \mathbb{E}\!\left[\mathcal{M}(\mathbf{w})\mathcal{M}(\mathbf{w})^{\!\top}\right], \nonumber\\
    [\boldsymbol{\Sigma}_Y]_{ij} &= R_Y(iT_s-jT_s),
    \label{eq:SigmaY_def}
\end{align}
where $R_Y(\tau)=\mathbb{E}[\mathcal{M}(w(t))\mathcal{M}(w(t+\tau))]$ and the mean is zero because $\mathcal{M}(\cdot)$ is odd and $w$ is symmetric.

Since $\bfw$ is correlated through $R_w(\tau)=\sigma_{\mathrm{lpf}}^2\rho(\tau)$, $\rho(\tau)=\sinc(2W\tau)$, we need the joint second-order statistics of $\Mod{\bfw}$, i.e.\ the matrix
\begin{align}
    \SigY &\triangleq \E\!\left[\Mod{\bfw}\Mod{\bfw}^{\!\top}\right], \nonumber\\
    [\SigY]_{ij}&=R_Y(iT_s-jT_s),
    \label{eq:SigmaY_def}
\end{align}
where $R_Y(\tau)=\E[\Mod{w(t)}\Mod{w(t+\tau)}]$ and the mean is zero because $\Mod{\cdot}$ is odd and $w$ is symmetric. Writing $\varphi(x)=\Mod{x}$ as a $2\lambda$-periodic sawtooth, its Fourier series is
\begin{equation}
    \varphi(x)=\sum_{\substack{n=-\infty\\n\ne0}}^{\infty} c_n e^{jn\pi x/\lambda}, \qquad c_n=\frac{j\lambda}{n\pi}(-1)^n.
    \label{eq:fourier_g}
\end{equation}
The coefficients come from integrating by parts over one period,
\begin{equation}
    c_n=\frac{1}{2\lambda}\int_{-\lambda}^{\lambda} x\,e^{-jn\pi x/\lambda}\,\mathrm{d}x
    =\frac{1}{2\lambda}\cdot\frac{2\lambda\cos(n\pi)}{-jn\pi/\lambda}
    =\frac{j\lambda}{n\pi}(-1)^n,
    \label{eq:cn_integral}
\end{equation}
since the remaining integral of $e^{-jn\pi x/\lambda}$ over a full period is zero. We also have $c_0=0$ because $\varphi$ is odd.

Let $w_1=w(t)$, $w_2=w(t+\tau)$ be jointly Gaussian with variance $\sigma_{\mathrm{lpf}}^2$ and correlation $\rho(\tau)$. Substituting~\eqref{eq:fourier_g} into $R_Y(\tau)\triangleq\E[\varphi(w_1)\varphi(w_2)]$ and taking the expectation inside the sums gives
\begin{equation}
    R_Y(\tau)=\sum_{n\ne0}\sum_{m\ne0} c_n c_m\,
    \E\!\left[e^{j\frac{\pi}{\lambda}(nw_1+mw_2)}\right].
    \label{eq:double_sum}
\end{equation}
Each expectation is the joint characteristic function of a bivariate Gaussian,
\begin{equation}
    \E\!\left[e^{j\frac{\pi}{\lambda}(nw_1+mw_2)}\right] = e^{-\alpha(n^2+m^2)}\,e^{-2\alpha nm\,\rho(\tau)},
    \label{eq:JCF}
\end{equation}
with $\alpha\triangleq\sigma_{\mathrm{lpf}}^2\pi^2/(2\lambda^2)$. The first factor depends only on the marginal variances, so the whole dependence on $\tau$ sits in the second factor. Taylor-expanding that factor in powers of $\rho(\tau)$ and inserting it into~\eqref{eq:double_sum} yields
\begin{equation}
    R_Y(\tau)=\sum_{k=0}^{\infty}\frac{\left(-2\alpha\rho(\tau)\right)^k}{k!}
    \sum_{n\ne0}\sum_{m\ne0} c_n c_m\,(nm)^k\, e^{-\alpha(n^2+m^2)}.
    \label{eq:expand}
\end{equation}
Because $(nm)^k$ factorizes, the inner double sum is a product of two identical single sums:
\begin{equation}
    R_Y(\tau)=-\frac{\lambda^2}{\pi^2}\sum_{k=0}^{\infty}\frac{\left(-2\alpha\rho(\tau)\right)^k}{k!}
    \left(\sum_{n\ne0}(-1)^n n^{k-1}e^{-\alpha n^2}\right)^{\!2}.
    \label{eq:Sk}
\end{equation}

Replacing $n$ with $-n$ in the inner sum leaves both $(-1)^n$ and $e^{-\alpha n^2}$ invariant, whereas the factor $n^{k-1}$ acquires a sign change of $(-1)^{k-1}$. Consequently, for even $k$, the summand is odd with respect to $n$, causing the symmetric sum over $n \neq 0$ to vanish. For $k=0$, this agrees with the zero-mean property of $\varphi(w)$. Thus, non-vanishing terms exist only for odd integers $k = 2m+1$. Restricting the summation domain to $n \ge 1$ by symmetry yields $B_m$, as defined in~\eqref{eq:Acoeffs}. Since $(-2\alpha\rho)^{2m+1} = -(2\alpha)^{2m+1}\rho^{2m+1}$, this resulting negative sign cancels the leading minus sign in~\eqref{eq:Sk}, ensuring that only \emph{odd} powers of $\rho(\tau)$ persist:
\begin{equation}
    R_Y(\tau) = \sum_{m=0}^{\infty} A_{2m+1}\,[\rho(\tau)]^{2m+1},
    \label{eq:RY}
\end{equation}
\begin{align}
    A_{2m+1} &= \frac{\lambda^2(2\alpha)^{2m+1}}{\pi^2(2m+1)!}B_m^2, \nonumber\\
    B_m &= 2\sum_{n=1}^{\infty}(-1)^n n^{2m} e^{-\alpha n^2}.
    \label{eq:Acoeffs}
\end{align}
 Above $\lambda/\sigma_{\mathrm{lpf}}\geq3$, the first-order term therefore accounts for almost all of $R_Y$, and $R_Y(\tau)/R_Y(0)$ is practically indistinguishable from $\rho(\tau)$. This is visible in Figs.~\ref{fig:autocorr}(a) and~(b), where the two curves overlap over the whole lag range. Below $\lambda/\sigma_{\mathrm{lpf}}=3$, the higher odd powers $\rho^3,\rho^5,\dots$ start to carry a significant share of the correlation. Since these powers have a narrower main lobe and much weaker sidelobes than $\rho$, the folded noise decorrelates faster than the unfolded noise. At $\lambda/\sigma_{\mathrm{lpf}}=2$ (Fig.~\ref{fig:autocorr}(c)) the main lobe of $R_Y(\tau)/R_Y(0)$ is visibly narrower.

\section{Quantization Noise: The Inflated-Covariance Approximation}
\label{app:quant}

This appendix proves the claim used in Section~\ref{sec:likelihood_truncation}: that the convolution $(\phi_{\SigW}*f_{\bfq})(\cdot)$ of the channel-noise density $\phi_{\SigW}(\cdot)$ with the uniform quantization density $f_{\bfq}(\cdot)$ equals, up to $O(\sigma_q^4)$, the density $\phi_{\SigTot}(\cdot)$ of a single Gaussian with the inflated covariance $\SigTot=\SigW+\sigma_q^2\bfI$. 

\subsection{Characteristic functions of the noise components}
Write $\bfn\triangleq\bfw+\bfq$. The density we must approximate is $p_{\bfn}(\bfx)=(\phi_{\SigW}*f_{\bfq})(\bfx)$, the density of the sum of the independent channel noise $\bfw$ and quantization noise $\bfq$. Because $\bfw$ and $\bfq$ are independent, the characteristic function of their sum is the product of their characteristic functions,
\begin{equation}
    \Phi_{\bfn}(\bfom) \triangleq \E\big[e^{j\bfom^\top\bfn}\big] = \Phi_{\bfw}(\bfom)\,\Phi_{\bfq}(\bfom).
    \label{eq:cf_product}
\end{equation}
Since $\bfw\sim\mathcal N(\mathbf 0,\SigW)$, its characteristic function is the standard Gaussian one,
\begin{equation}
    \Phi_{\bfw}(\bfom) = \exp\!\Big(-\tfrac12\bfom^\top\SigW\bfom\Big).
    \label{eq:cf_w}
\end{equation}
For the quantization noise, each $q_n\sim\mathcal U(-\Delta/2,\Delta/2)$ is independent, so its characteristic function is the exact product
\begin{equation}
    \Phi_{\bfq}(\bfom) = \prod_{n=1}^N \frac{\sin(\omega_n\Delta/2)}{\omega_n\Delta/2}.
    \label{eq:cf_q_exact}
\end{equation}
Using $\sin(u)/u=1-u^2/6+O(u^4)$ and $\sigma_q^2=\Delta^2/12$, each factor expands as
\begin{align}
    \frac{\sin(\omega_n\Delta/2)}{\omega_n\Delta/2} &= 1-\frac{\omega_n^2\Delta^2}{24}+O(\Delta^4) \nonumber\\
    &= 1-\frac{\sigma_q^2}{2}\omega_n^2+O(\sigma_q^4).
\end{align}
Multiplying the $N$ factors ,the cross terms between different $n$ are themselves $O(\sigma_q^4)$ and do not contribute at this order, gives
\begin{equation}
    \Phi_{\bfq}(\bfom) = 1-\frac{\sigma_q^2}{2}\|\bfom\|^2+O(\sigma_q^4).
    \label{eq:cf_q_expanded}
\end{equation}

\subsection{Matching to a single inflated Gaussian}
Substituting~\eqref{eq:cf_w} and~\eqref{eq:cf_q_expanded} into~\eqref{eq:cf_product} gives
\begin{align}
    \Phi_{\bfn}(\bfom) ={}& \exp\!\Big(-\tfrac12\bfom^\top\SigW\bfom\Big) \nonumber\\
    &\times\bigg[1-\frac{\sigma_q^2}{2}\|\bfom\|^2\bigg]+O(\sigma_q^4).
    \label{eq:cf_n}
\end{align}
Now consider a single Gaussian $\bfm\sim\mathcal N(\mathbf 0,\SigTot)$ with $\SigTot=\SigW+\sigma_q^2\bfI$. Because $\bfom^\top(\sigma_q^2\bfI)\bfom=\sigma_q^2\|\bfom\|^2$, its characteristic function factors exactly as
\begin{align}
    \Phi_{\bfm}(\bfom) ={}& \exp\!\Big(-\tfrac12\bfom^\top\SigTot\bfom\Big) \nonumber\\
    ={}& \exp\!\Big(-\tfrac12\bfom^\top\SigW\bfom\Big)\exp\!\Big(-\tfrac{\sigma_q^2}{2}\|\bfom\|^2\Big).
    \label{eq:cf_gaussian}
\end{align}
Expanding the second exponential exactly as in~\eqref{eq:cf_q_expanded},
\begin{equation}
    \exp\!\Big(-\tfrac{\sigma_q^2}{2}\|\bfom\|^2\Big) = 1-\frac{\sigma_q^2}{2}\|\bfom\|^2+O(\sigma_q^4),
\end{equation}
so that, comparing with~\eqref{eq:cf_n},
\begin{equation}
    \Phi_{\bfm}(\bfom) = \Phi_{\bfn}(\bfom) + O(\sigma_q^4)
    \label{eq:cf_match}
\end{equation}
for every $\bfom$: the two characteristic functions agree to second order in $\sigma_q$. Since a characteristic function determines its density uniquely, through the (inverse) Fourier transform, this proves
\begin{equation}
    p_{\bfn}(\bfx) = \phi_{\SigTot}(\bfx) + O(\sigma_q^4), \qquad \SigTot=\SigW+\sigma_q^2\bfI,
    \label{eq:quant_result}
\end{equation}
recovering~\eqref{eq:SigmaTotal}. Substituting~\eqref{eq:quant_result} into the post-quantization likelihood $(\phi_{\SigW}*f_{\bfq})(\bfr(\bfa))$ of Section~\ref{sec:likelihood_truncation} and taking logarithms reproduces the Mahalanobis rule~\eqref{eq:mahalanobis_ml_quantized}.

\subsection{Physical interpretation}
Equation~\eqref{eq:quant_result} captures the precise sense in which quantization noise increases the channel noise variance. Specifically, the characteristic function of a sum of independent random vectors equals the product of their individual characteristic functions. For two independent Gaussian vectors, this product is exact and corresponds to the direct summation of their covariance matrices. Although $\bfq$ is uniform rather than Gaussian, its characteristic function matches that of a Gaussian surrogate up to second order in $\bfom$~\eqref{eq:cf_q_expanded}, sharing the same leading coefficient $\sigma_q^2$. Consequently, the discrepancy between $\bfq$ and a Gaussian noise model arises only at $O(\sigma_q^4)$. This approximation improves as $\sigma_q \ll \sigma_{\mathrm{lpf}}$, i.e., as the quantization step $\Delta$ becomes small relative to the channel-noise standard deviation, which justifies its use in Sections~\ref{sec:likelihood_truncation} and~\ref{subsec:block_search} for $R \ge 1$ bits and $\lambda/\sigma_{\mathrm{lpf}} \ge 3$.

\bibliographystyle{IEEEtran}
\bibliography{refs_latest}

\end{document}